# Spin rotation of the planar channeled antiproton in bent crystal

Yu.P. Kunashenko[1]

*Physics and Mathematics Faculty, Tomsk State Pedagogical University, Kievskaya Str., 60, 634061 Tomsk, Russia*

In the present paper, we derive the exact solution of the BMT equation for the spin rotation of relativistic negatively charged particles passing through a thin bent crystal in the regime of planar channeling. It is shown that during the passage through the crystal a particle polarization vector oscillates. If one takes into account, the energy and angular divergence of the initial particle beam these oscillations die fast with particle penetration length into a crystal, and particle spin rotation is determined by the crystal bending angle.

## I. INTRODUCTION

The spin of the elementary particles is one of the fundamental differences between the elementary particles and macroscopic bodies. Therefore, investigation of the problems connected with spin effects is undoubted of great interest. One of the most important and interesting questions is the interaction of the particle spin with the external electromagnetic field.

In the paper of the Bargmann, Michel, and Teledgi [1] it was obtained quasi-classical equation (BMT equation) describing the spin motion accompanying the particle's traveling in an external electromagnetic field. The connection between spin rotation angle and curvature of the trajectory of a relativistic charged particle moving in the electromagnetic field was established in the paper [2].

In 1976, Tsyganov [3] shown that high-energy charged particles in a bent crystal may move in the channeling regime, so passing along a curved trajectory and deviating from the initial direction. Now, the experiments on channeling in bent crystals are performed at the largest world's accelerator centers [4–9].

For the first time, Baryshevsky suggested the new possibility of investigation of the particle spin interaction with the electromagnetic field of crystal in [10]. In this paper, it was predicted rotation of spin of relativistic positively charged particles passing through the bend and straight crystals in the planar channeling regime. Further development of theory and methods of modeling was performed in [11-17]. Experiments, which confirmed the existence of the existence spin rotation effect, were realized in Fermilab [18-20].

---

[1] *E-mail addresses:* `kunashenko@tspu.edu.ru`

The phenomenon of the particle spin rotation in the bent crystal now is planned to use for the measurement of magnetic moment, electric and magnetic dipole moment of short-lived particles [11-17, 21-25].

For a detailed description and planning of experiments on the spin rotation of a charged particle in a bent crystal, the method of computer simulation has proven itself well [14-16]. Nevertheless, from the point, of understanding and clarify the physics of the process it is very important to have an analytical solution. Spin rotation in the different electromagnetic fields has been investigated in some theoretical papers. For example, in [26] (and references therein) the dynamics of the spin of a charged particle with an anomalous magnetic moment in an arbitrary constant magnetic field are investigated. A class of fields is found in which the solution of the BMT equation can be presented in an analytical form.

A new method of calculation of the negatively charged channeled particle spin rotation angle in straight line crystal was suggested in [27, 28]. The main idea of this method is as follows: At the first step, the particle trajectory in the continuous potential of the crystal planes is calculated. Then, using the obtained trajectory, it is found the exact solution of the BMT equation.

There are other possibilities for the interaction of the spin with the electromagnetic field. For an example, the neutron is a neutral particle, but, as it was shown by Schwinger [29], it can be scattered by an atom's electrical field due to the presence of spin (and, therefore, of the magnetic moment). Schwinger neutron scattering was discovered in [30].

Similarly, to the coherent effect on the scattering of fast charged particles in oriented crystals, one can expect the emergence of this effect for neutrons in crystals, in addition to regular scattering by the nucleus. For the first time, the existence of coherent effects in Schwinger neutron scattering was shown in [31, 32]. The only one experiment concerned with measuring the cross-section of Schwinger neutron scattering in a germanium crystal was carried out in [33]. More detailed calculations of the coherent Schwinger scattering of fast neutrons in a crystal and determinations of the optimal conditions for experimental studies of this phenomenon were done in [34-36].

Another mechanism of neutron interaction with the electromagnetic field is the emission of photons by neutrons. For the first time, radiation from neutrons in an external electromagnetic field has been theoretically studied in [37–39]. This new type of radiation produced at the interaction of the anomalous magnetic moment and magnetic field was named ''spin'' light [39]. The theory of coherent bremsstrahlung from neutrons in the crystal was derived in [40, 41].

The interaction of charged particles with oriented particles for a long time is a perceptive field of research both from the point of view of theory and experiment. There are coherent processes and channeling phenomena. The passage of particles through crystals is accompanied by various physical phenomena, for example, coherent bremsstrahlung and coherent pair production, channeling radiation, the electron-positron pair creation, and others. These phenomena are described in detail in a number of monographs and original articles (see for example [42-63] and references therein).

In the present paper, we derive the exact solution of the BMT equation for investigation of the rotation of spin of relativistic negative charged particles passing through a bent crystal. The case of positively charged particles was detailed considered in the paper [3]. Using the obtained solutions of equations, we particularly investigate the spin rotation of relativistic antiproton channeled in the tungsten crystal.

## II. PLANAR CHANNELED PARTICLE'S TRAJECTORY IN THE BENT CRYSTAL

In order to find a solution to the BMT equation one has to know the particle trajectory in the external electromagnetic field. Therefore, we start our consideration with the discussion of the relativistic charged particle passage through the bent crystal in the regime of planar channeling. We represent the crystal planes in the form of short arcs of concentric circles.

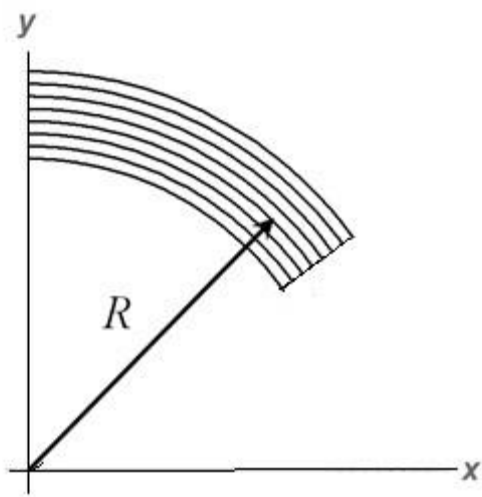
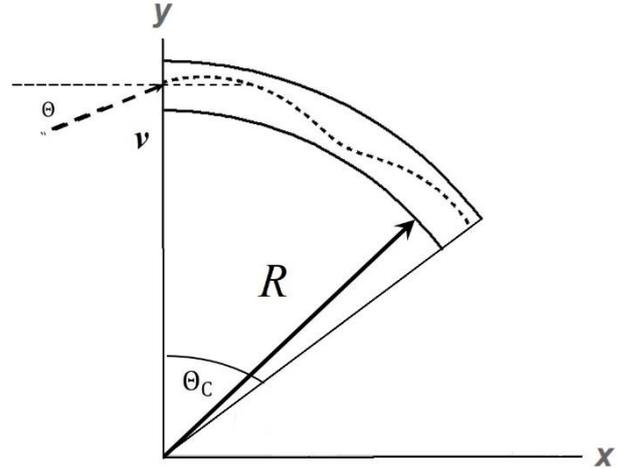

FIG. 1a. The bend crystal is shown schematically: arcs are crystal planes; arrow R is the radius of curvature of the crystal.

FIG. 1b. It is shown particle trajectory in the single channel of crystal. Solid lines indicate channel; the dashed line is particle trajectory; solid arrow R is the radius of curvature of the plane; the dashed arrow $v$ is the vector of initial particle velocity and angle $\Theta$ is the particle angle of entry with respect to the crystal plane. The angle $\Theta_C$ is the crystal bent angle.

In Fig.1a the bent crystal is shown schematically: arcs are crystal planes; arrow R is the radius of curvature of the crystal. Fig. 1b shows the particle trajectory in the single channel of crystal.

Since we are interested in the motion in the channeled state (finite motion along to a separate crystal plane), it is possible to use the continuous potential of a single plane. One of the simplest and reasonably models of the crystal plane potential is the modified Pöschl-Teller potential (see for example [46,51,62] and references therein)

$$V(x) = -V_0 cosh^{-2}\left(\frac{x}{b}\right). \qquad (1)$$

Here $V_0$ is the depth of the potential well and $b$ is constant. In order to take into account, the curvature of the crystal, we rewrite the potential in polar coordinates as follows

$$V_B(\rho) = V_0 - V(\rho) = V_0\left[1 - cosh^{-2}\left(\frac{r}{b}\right)\right]. \qquad (2)$$

Now $r$ is particle position inside the channel. If the $\rho$ is particle radial coordinate and $R$ is the curvature radius of the crystal plane creating the potential $V(\rho)$ then $r = \rho - R$. When we study particle motion in the separate channel then with very good accuracy, we can assume that R is equal to the radius of curvature of the crystal. For the convenience of further calculations, we have shifted the potential energy by $V_0$.

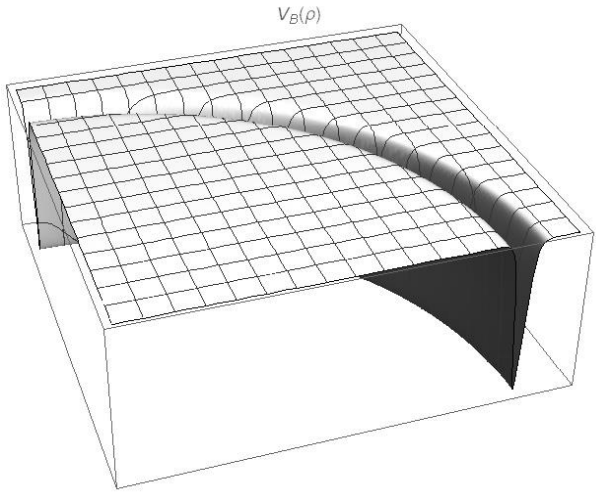

FIG.1c. The potential of a single plane (2) of the bent crystal.

It is convenient to choose the coordinate system as it follows: The X-axis is parallel to the crystal plane for straight-line crystal and, the Y-axis is perpendicular to this plane. It means, the initial longitudinal velocity ($v_\parallel$) is directed along the X-axis, whereas the initial transverse velocity ($v_\perp$) of the particle is parallel to the Y-axis, the Z-axis is perpendicular to the plane of motion.

The angular particle momentum $\mathbf{M} = [\mathbf{r}, \mathbf{p}]$ is directed parallel to Z-axis and equals

$$M = \gamma m \rho^2 \dot{\varphi} \approx \gamma m R^2 \dot{\varphi}. \qquad (3)$$

With $\dot{\varphi}$ being the particle angular velocity. It should be noted that in the initial time moment one can write $M = \gamma m v_\parallel \rho \approx \gamma m v_\parallel R$.

The relativistic energy of the channeled particle is equal to

$$E = c\sqrt{m c^2 + p^2} + V_B(\rho). \tag{4}$$

It is convenient to represent the particle total energy E as a sum of longitudinal energy $E_{||}$ and transverse one $\varepsilon_\perp$

$$E = E_{||} + \varepsilon_\perp.$$

Substitution E and $V_B(\rho)$ in the Eq. (4), squaring the result, we get

$$\left[(E_{||} + \varepsilon_\perp) - (V_0 - V(\rho))\right]^2 = m^2c^4 + c^2p^2. \tag{5}$$

Considering that for channeled particle the condition $E_{||} \gg \varepsilon_\perp, V_0, V(\rho)$ is satisfied and keeping the first-order terms of smallness on the left side of the equation (5) we arrive at

$$E_{||}^2 + 2E_{||}\varepsilon_\perp - 2E_{||}V_0 + 2E_{||}V(\rho) = m^2c^4 + c^2p^2. \tag{6}$$

In the polar coordinates particle momentum squared is

$$p^2 = p_\rho^2 + p_\varphi^2 = \gamma^2 m^2 \dot\rho^2 + \gamma^2 m^2 \rho^2 \dot\varphi^2. \tag{7}$$

The angular velocity according to Eq. (3) is equal to $\dot\varphi = \frac{M}{\gamma m \rho^2}$, after substitution $\dot\varphi$ into Eq. (7) we find

$$p^2 = \gamma^2 m^2 \dot\rho^2 + \frac{M^2}{\rho^2}.$$

Now the Eq. (5) takes the form

$$\left(E_{||}^2 - \frac{M^2 c^2}{\rho^2} - m^2 c^4\right) + 2E_{||}(\varepsilon_\perp - V_0 + V(\rho)) = \gamma^2 m^2 c^2 \dot\rho^2.$$

If we take into account that in the initial time moment $M = \gamma m v_{||} \rho$ and $E_{||} = \gamma m c^2$, $\gamma = (1 - v_{||}^2/c^2)^{-1/2}$ then we obtain

$$E_{||}^2 - \frac{M^2 c^2}{\rho^2} - m^2 c^4 = 0,$$

and we arrive at

$$2E_{||}(\varepsilon_\perp - V_0 + V(\rho)) = \gamma^2 m^2 c^2 \dot\rho^2. \tag{8}$$

Solving the last equation for $\dot\rho^2$ we finally find

$$\dot\rho^2 = \left(\frac{d\rho}{dt}\right)^2 = \frac{2}{\gamma m}(\varepsilon_\perp - V_0 + V(\rho)). \tag{9}$$

Eq. (9) similar to the equation describing the transverse motion of a negatively charged particle channeling in a straight crystal (if the $\rho$ is replaced by x). As it follows from Eq. (9) the radial coordinate $\rho$ is determined by the expression

$$t = \int \frac{d\rho}{\sqrt{\frac{2}{\gamma m}(\varepsilon_\perp - V_0 + V(\rho))}}. \tag{10}$$

Integration of Eq. (10) for potential (2) can be carried out analytically, the result of integration is

$$\rho = b\, Arcsin\left[A \sin\frac{t-t_0}{\tau}\right] + R. \tag{11}$$

Constant of integration $t_o$ determined by initial conditions. Other notations are $\tau = \sqrt{\frac{\gamma m b^2}{2(V_0-\varepsilon_\perp)}}$, $A = \sqrt{\frac{\varepsilon_\perp}{(V_0-\varepsilon_\perp)}}$.

The angel φ we find from condition $M = \gamma m \rho^2 \dot{\varphi} \approx \gamma m R^2 \dot{\varphi}$

$$\varphi = \Omega t, \quad \Omega = \frac{M}{\gamma m R^2}. \tag{12}$$

### III. THE SOLUTION OF THE BMT EQUATION

Let us consider the spin rotation of the planar channeled negatively charged particle in the bent crystal. The BMT equation has a form [1]

$$\frac{d\boldsymbol{\zeta}}{dt} = [\boldsymbol{Z}, \boldsymbol{\zeta}]. \tag{13}$$

Here *m* is the particle rest mass, **ζ** is the polarization vector (unit vector directed like spin) $\boldsymbol{\zeta} = (\zeta_x, \zeta_y, \zeta_z)$, and vector **Z** is

$$\boldsymbol{Z} = -\frac{e}{2mc} g \left\{ \boldsymbol{H} - \frac{\gamma-1}{\gamma} \boldsymbol{l}(\boldsymbol{H}\boldsymbol{l}) + \left[\boldsymbol{E}, \frac{\boldsymbol{v}}{c}\right] \right\} - (\gamma - 1) \left[\boldsymbol{l}, \frac{d\boldsymbol{l}}{dt}\right]. \tag{14}$$

Here γ is the particle relativistic factor; *l* is the unit vector in the direction of particle speed, *g* is the gyromagnetic ratio (the magnetic moment **μ**=$\frac{eg}{2mc}\hbar$**s**, with **s** being the spin vector of the particle), *e* is the elementary charge, *v* is channeled particle velocity vector, *c* is the speed of light in vacuum, **E** and **H** are the electric field strength vector and magnetic field vector at the point of particle location.

In a crystal, the magnetic field is absent **H** = 0 and, consequently, the vector **Z** has the form

$$\boldsymbol{Z} = -\frac{e}{2mc} g \left[\boldsymbol{E}, \frac{\boldsymbol{v}}{c}\right] - (\gamma - 1) \left[\boldsymbol{l}, \frac{d\boldsymbol{l}}{dt}\right]. \tag{15}$$

The channeled particle's trajectory in the bent crystal has been found in the polar coordinates; therefore, we should rewrite the BMT equation in the polar coordinates too.

The electric field strength of the bent crystal plane continuous potential $V_B(\rho)$ (2) in the polar coordinates has only one component $E_\rho$, which is equal to

$$E_\rho = -\frac{2V_0 \operatorname{sech}\left(\frac{r}{b}\right)^2 \operatorname{th}\left(\frac{r}{b}\right)}{b}.$$

Substituting electric field strength and particle trajectory into BMT equation after some algebra, we find

$$\begin{cases} \dfrac{d\zeta_\rho(t)}{dt} = \Lambda\, \zeta_\varphi(t) \\ \dfrac{d\zeta_\varphi(t)}{dt} = -\Lambda\, \zeta_\rho(t) \\ \dfrac{d\zeta_z(t)}{dt} = 0 \end{cases} ; \qquad (16)$$

here we denote

$$\Lambda = \frac{AR\Omega \sin\left(\frac{t-t_0}{\tau}\right)}{\left(1+A^2 \sin^2\left(\frac{t-t_0}{\tau}\right)\right)^{1/2}} \left[ -\frac{egV_0}{b\,mc^2\left(1+A^2 \sin^2\left(\frac{t-t_0}{\tau}\right)\right)} + \frac{(1+A^2)b(\gamma-1)}{A^2 b^2 \cos\left(\frac{t-t_0}{\tau}\right)+R^2\tau^2\Omega^2\left(1+A^2 \sin^2\left(\frac{t-t_0}{\tau}\right)\right)} \right].$$

The equality $\dfrac{d\zeta_z(t)}{dt} = 0$ means that the component of the polarization vector $\zeta_z$ is conserved. Therefore, without loss of generality, we choose the vector $\boldsymbol{\zeta}$ in the following form $\boldsymbol{\zeta} = (\zeta_\rho, \zeta_\varphi, 0)$.

The solutions of the system of equations (16) for the initial conditions $\xi_\rho(0) = 1$ and $\xi_\varphi(0) = 0$ are

$$\begin{cases} \zeta_\rho(t) = \cos(\Psi(t)) \\ \zeta_\varphi(t) = \sin(\Psi(t)) \end{cases}. \qquad (17)$$

Here the following notations are introduced

$$\Psi(t) = C\left[\Theta\left(\frac{t-t_o}{\tau}\right) - \Theta\left(\frac{t_o}{\tau}\right)\right] + (\gamma - 1)\left[\Phi\left(\frac{t-t_o}{\tau}\right) - \Phi\left(\frac{t_o}{\tau}\right)\right],$$

and

$$\Theta(x) = \frac{\cos(x)}{\sqrt{2+A^2-A^2\cos(2x)}}, \quad C = \frac{\sqrt{2}\,A\,R\,\tau\,\Omega\,egV_0}{\gamma mc^2\,(1+A^2)b},$$

$$\Phi(x) = \text{Arcctg}\left(\frac{R\tau\Omega\sqrt{2+A^2-A^2\cos(2x)}\,\sec(x)}{\sqrt{2}Ab}\right).$$

The choice $\xi_\rho(0) = 1$, $\xi_\varphi(0) = 0$ means that in Cartesian coordinates we have $\xi_x(0) = 0$, $\xi_y(0) = 1$ i.e. in the initial time moment particle spin is directed perpendicular to the crystal plane.

The obtained expressions (17) it is convenient to transform into Cartesian coordinates. For the coordinate system shown in Fig. 1a. the rules of transformation are as follows

$$\begin{cases} \zeta_x(t) = \zeta_\rho(t)\cos(\pi - \varphi(t)) - \zeta_\varphi(t)\sin(\pi - \varphi(t)) \\ \zeta_y(t) = \zeta_\rho(t)\sin(\pi - \varphi(t)) + \zeta_\varphi(t)\cos(\pi - \varphi(t)) \end{cases}.$$

After transformation, we have

$$\begin{cases} \zeta_x(t) = \sin(\varphi(t) + \Psi(t)) \\ \zeta_y(t) = \cos(\varphi(t) + \Psi(t)) \end{cases}. \qquad (13a)$$

If at the initial moment of time, the spin of the particle is directed parallel to the plane of the crystal ($\xi_x(0) = 1$, $\xi_y(0) = 0$), then the solutions from (13a) change places.

$$\begin{cases} \zeta_x(t) = \cos(\varphi(t) + \Psi(t)) \\ \zeta_y(t) = \sin(\varphi(t) + \Psi(t)) \end{cases} \quad (13b)$$

When we omit the angle $\varphi(t)$ in formula (13a), we obtain the equations for spin rotation in the straight crystal [19,20]

$$\begin{cases} \zeta_x(t) = \sin(\Psi(t)) \\ \zeta_y(t) = \cos(\Psi(t)) \end{cases} \quad (14)$$

For further analysis, it is convenient to introduce the spin rotation angle $\phi$

$$\phi = arcsin\left(\frac{\zeta_y(t)}{\sqrt{\zeta_x^2(t)+\zeta_y^2(t)}}\right) = arcsin\left(\zeta_y(t)\right), \quad (15)$$

here we take into account that $\zeta_x^2(t) + \zeta_y^2(t) = 1$.

The results of the calculation of the dependence of polarization vector components $\zeta_x(t), \zeta_y(t)$, and of spin rotation angle $\phi$ upon particle's penetration depth into crystal are shown in Fig2. The target is tungsten crystal oriented by plane (100) the relativistic factor of antiproton equals $\gamma=10^6$. Fig.2a shows polarization vector components $\zeta_x(t), \zeta_y(t)$ and Fig.2b shows spin rotation angle $\phi$. The initial angle of the velocity of the antiproton with respect to crystal plane $\Theta = 0.25\,\Theta_L$, here $\Theta_L = \sqrt{\frac{2V_0}{\gamma\,mc^2}}$ is the Lindhard critical angle for planar channeling $\Theta_L = 4.2\,10^{-7}$ rad, entry point relative to channel center $x_o = 0.45\,\dot{A}$, radius of curvature of the crystal *R=1 m.*

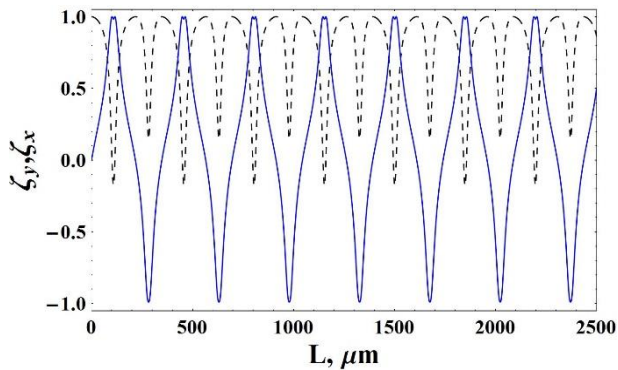 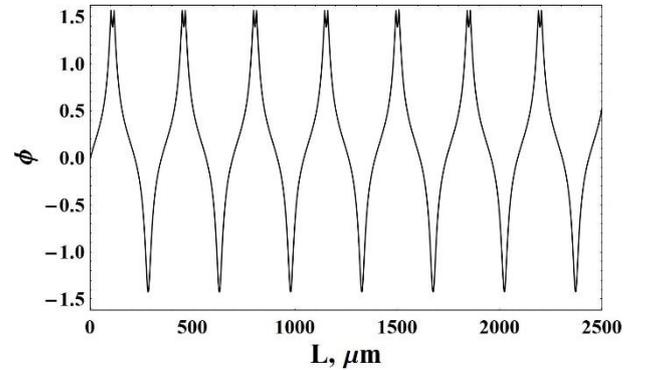

FIG.2a Polarization vector components $\zeta_x$ – solid line and $\zeta_y$ – dashed line as a function of particle's penetration depth into crystal L, $\Theta = 0.25\,\Theta_L$, $x_o = 0.29\,\dot{A}$.

FIG.2b Spin rotation angle $\phi$ as a function of particle's penetration depth into crystal L, $\Theta = 0.29\,\Theta_L$, $x_o = 0.21\,\dot{A}$.

Fig. 3 shows the same as Fig. 2 but for another entry point $x_o = 0.\dot{A}$; the other parameters as on Figs. 2.

It follows from Figs. 2 and Figs. 3 that the behavior the dependence of the polarization vector components on the depth of penetration of the antiproton into the crystal variate with changing the point of entry of the particle into the channel. On the other hand, it is impossible to fix the particle entry point into the crystal. Therefore, the polarization vector components $\zeta_x$ and $\zeta_y$ should be averaged over the entry points.

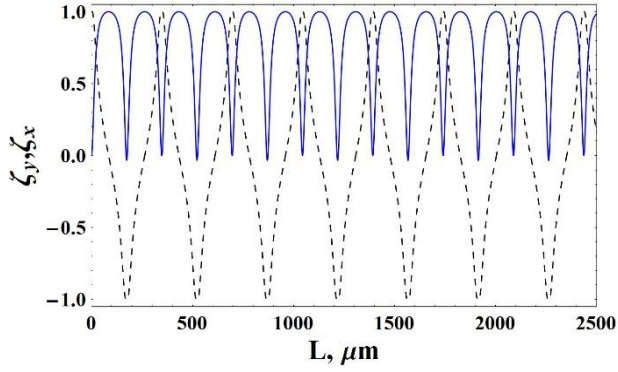 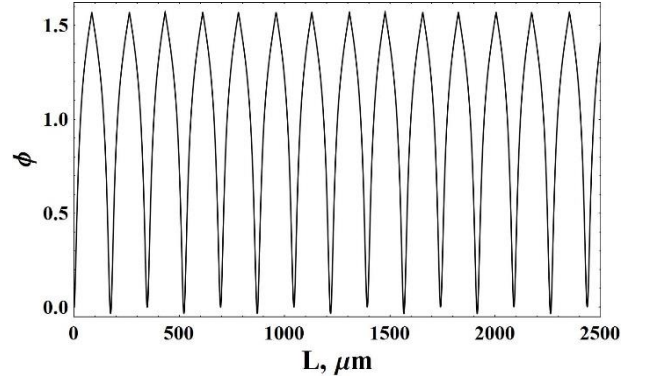

FIG.3a Polarization vector components $\zeta_x$, – solid line, $\zeta_y$ – dashed line as a function of particle's penetration depth into crystal $L$, $\Theta = 0.25\,\Theta_L$, $x_o = 0.\dot{A}$.

FIG.3b Spin rotation angle $\phi$ as a function of particle's penetration depth into crystal $L$, $\Theta = 0.25\,\Theta_L$, $x_o = 0.\dot{A}$.

In order to average, the polarization vector components $\zeta_x$ and $\zeta_y$ we take N particle's entry points into the channel with coordinates $x_n$. All points are separated by distance $d/N$. For every point, we calculate the polarization vector components $\zeta_x$ and $\zeta_y$ sum them up and divide by $N$.

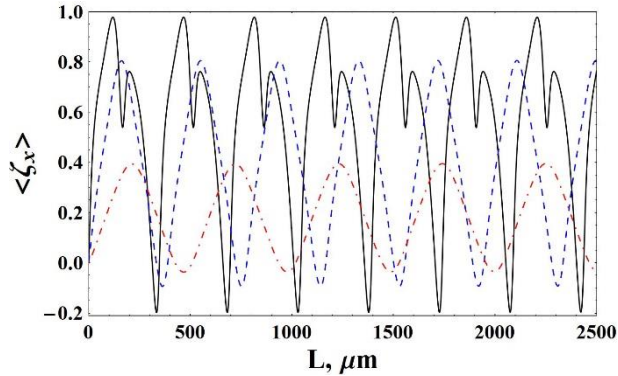 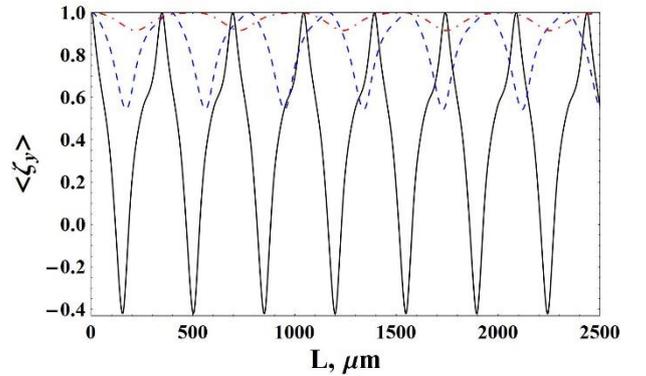

FIG. 4a. Dependence of the averaged polarization vector components $\langle\zeta_x\rangle$ as a function of particle's penetration depth into crystal $L$ for different initial antiproton angles with respect to crystal plane $\Theta =$

FIG. 4b. Dependence of the averaged polarization vector components $\langle\zeta_y\rangle$ as a function of particle's penetration depth into crystal $L$ for different initial antiproton angles with respect to crystal plane $\Theta =$

$$\langle \zeta_{x(y)}(t)\rangle = \frac{1}{N}\sum_{i=1}^{N}\zeta_{x(y)}(x_n,t) \tag{16a}$$

Using the found average values $\langle\zeta_x\rangle$ and $\langle\zeta_y\rangle$, we determine the average angle of spin rotation.

$$\langle\phi\rangle = arcsin\left(\frac{\langle\zeta_y(t)\rangle}{\sqrt{\langle\zeta_x(t)\rangle^2+\langle\zeta_y(t)\rangle^2}}\right), \tag{16b}$$

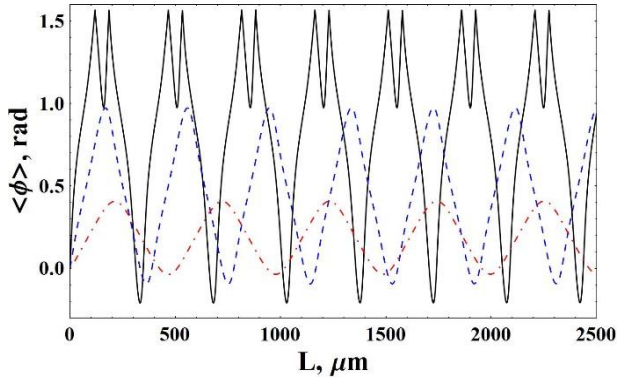

FIG. 4c. Dependence of the averaged spin rotation angle $\langle\phi\rangle$ as a function of particle's penetration depth into crystal $L$ for different initial antiproton angles with respect to crystal plane $\Theta = 0.25\,\Theta_L$ – solid line, $\Theta = 0.5\,\Theta_L$ – dashed line, $\Theta = 0.75\,\Theta_L$ – dotted line.

In Fig. 4 we plot dependence of the averaged polarization vector components $\langle\zeta_x(t)\rangle$ (Fig. 4a), $\langle\zeta_y(t)\rangle$ (Fig. 4b) and spin rotation averaged angle $\langle\phi\rangle$ (Fig. 4c) as a function of particle's penetration depth into crystal $L$ for different initial antiproton angles with respect to crystal plane $\Theta = 0.25\,\Theta_L$ – solid line, $\Theta = 0.5\,\Theta_L$ – dashed line, $\Theta = 0.75\,\Theta_L$ – dotted line. The target is tungsten crystal oriented by plane (100); the relativistic factor of antiproton equals $\gamma=10^6$.

From Fig. 4 follows, that averaged polarization vector components $\langle\zeta_x(t)\rangle$, $\langle\zeta_y(t)\rangle$ oscillates with particle's penetration depth into crystal $L$. The reason for this behavior is connected with the periodical trajectory of the planar channeled antiproton.

The contribution of the crystal curvature to the dependence of the components of the polarization vector and the rotation angle is very small. It can be estimated using the formula, which connects the spin rotation angle and particle trajectory curvature in the electromagnetic field [2]

$$\phi = \left((g-2)\frac{\gamma^2-1}{2\gamma}+\frac{\gamma-1}{\gamma}\right)\Psi, \tag{17}$$

with $\Psi$ being particle rotation angle in the external field. Putting $\Psi = \Omega t$ we get $\phi\sim 0.01$ rad for crystal thickness $L=1$ cm and $R=1$ m. The values of the crystal thickness L and the radius of curvature of the crystal R are chosen in accordance with the ones used in the experiments on charge particle channeling [4-9] and spin rotation in the bent crystal [18-20].

In order to demonstrate the influence of the crystal curvature on the spin rotation, we increase the magnitude of the angular velocity $\Omega$ by $10^3$ times. The results of calculations with increased angular velocity $\Omega_I = 10^3 \Omega$ are presented in Fig. 5 for the averaged polarization vector components $\langle \zeta_x(t) \rangle$ (Fig.5a) $\langle \zeta_y(t) \rangle$ (Fig. 5b) and average spin rotation angle $\langle \phi \rangle$ (Fig. 5c). To stress the influence of the crystal curvature, we additionally plot the results of the calculation for a straight crystal (dashed line). The target is tungsten crystal oriented by plane (100); the relativistic factor of antiproton equals $\gamma=10^6$, the initial angle of the antiproton with respect to crystal plane $\Theta = 0.5\,\Theta_L$.

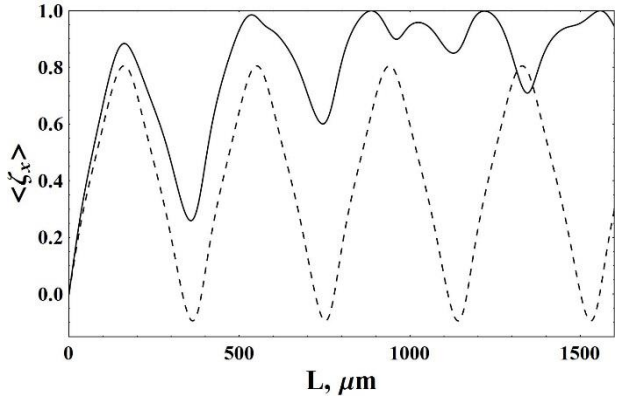
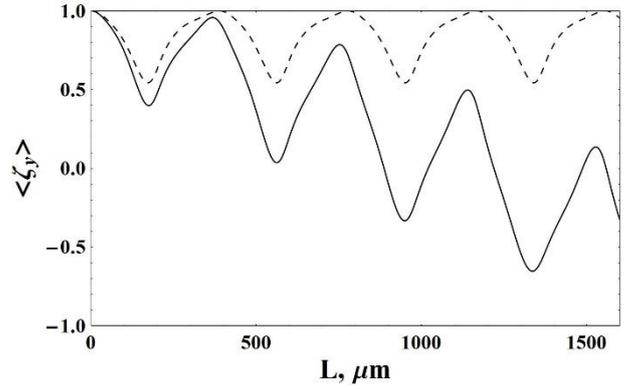

FIG. 5a Dependence of the averaged polarization vector components $\langle \zeta_x(t) \rangle$ as a function of particle's penetration depth into crystal L calculated using $\Omega_I$ – solid line, the result of calculation for a straight crystal - dashed line.

FIG. 5b Dependence of the averaged polarization vector components $\langle \zeta_y(t) \rangle$ as a function of particle's penetration depth into crystal L calculated using $\Omega_I$ – solid line, the result of calculation for a straight crystal - dashed line.

From Fig. 5 it follows that in the case of a greater angular velocity $\Omega_I$ additional rotation of the particle trajectory due to the curvature of the crystal leads to a significant change in the behavior of the dependence of the components of the polarization vector and the angle of rotation. The oscillation period and the shape of the curve change, the amplitude of spin rotation angle oscillations increases. These changes are increasing with the increasing of the particle's penetration depth into the crystal. The reason for this is that with increasing particle penetration depth into the crystal, the difference between the trajectory bending angles in a straight and bend crystal increases significantly in comparison in comparison with the case when antiproton has ordinary angular velocity Ω.

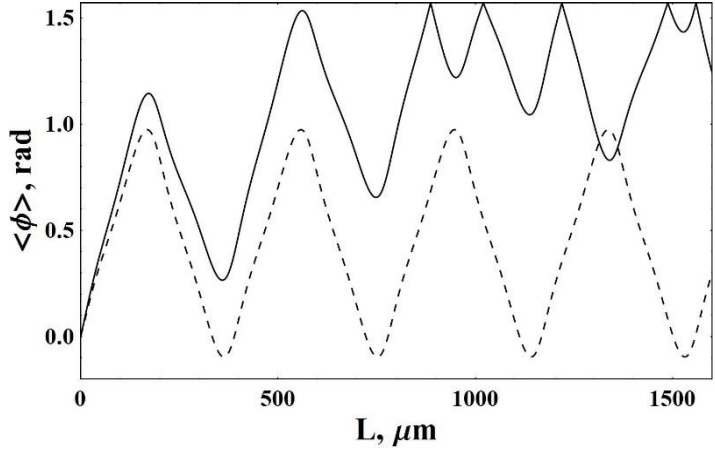

FIG. 5c Dependence of averaged spin rotation angle as ⟨ϕ⟩ a function of particle's penetration depth into crystal $L$ calculated using $\Omega_I$ – solid line, the result of calculation for a straight crystal - dashed line.

## IV.   INFLUENCE OF ENERGY AND ANGULAR DIVERGENCE

In real experiments, the antiproton beam is characterized by energy and angular divergence; therefore, we average the obtained results over the angular and energy spread.

To take into account the divergence, we used the Gaussian distribution

$$f(x) = \frac{1}{\sigma\sqrt{2\pi}} e^{-\frac{(x-x_0)^2}{2\sigma^2}}. \tag{18}$$

Here $x$ is the value of a distributed quantity that in our case is an antiproton's angle of entry into the crystal or antiproton's relativistic factor, with $x_0$ being mean and $\sigma$ being the standard deviation of the considered quantity.

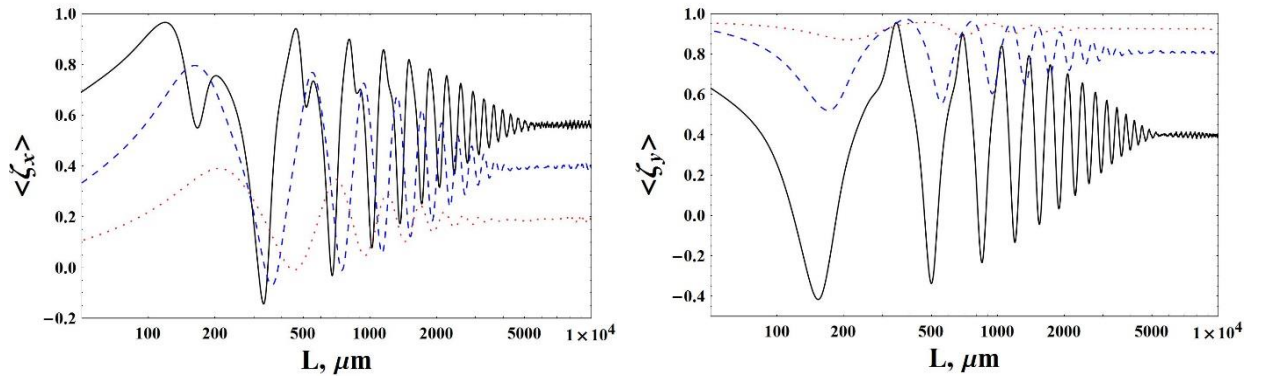

FIG. 6a Dependence of the averaged polarization vector components $\zeta_x(t)$ as a function of particle's penetration depth into crystal $L$ for different initial antiproton angles with respect to crystal plane $\Theta = 0.25\,\Theta_L$ – solid line, $\Theta = 0.5\,\Theta_L$ – dashed line, $\Theta = 0.75\,\Theta_L$ – dotted line, $\gamma_0 = 10^6$, $\sigma_\gamma = 0.05\,\gamma_0$, $\sigma_\gamma = 0.05\,\Theta_0$.

FIG. 6b Dependence of the averaged polarization vector components $\zeta_y(t)$ as a function of particle's penetration depth into crystal $L$ for different initial antiproton angles with respect to crystal plane $\Theta = 0.25\,\Theta_L$ – solid line, $\Theta = 0.5\,\Theta_L$ – dashed line, $\Theta = 0.75\,\Theta_L$ – dotted line, $\gamma_0 = 10^6$, $\sigma_\gamma = 0.05\,\gamma_0$, $\sigma_\gamma = 0.05\,\Theta$.

To take into account the energy and angular spreads in the antiproton beam, we additionally calculate according to the distribution (18) the entry angle and the energy (relativistic factor) for antiproton in every initial point. The further procedure is similar to the averaging excluding the angular and energy spreads (16a, 16b).

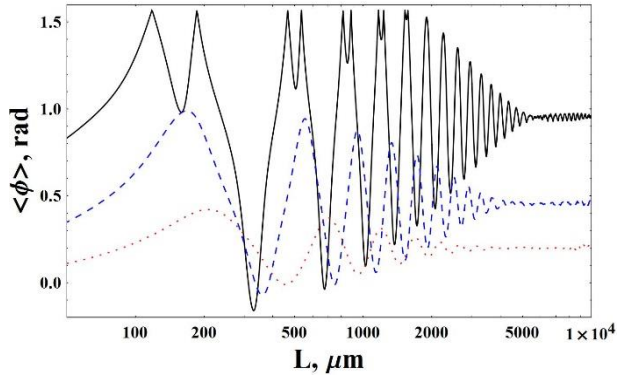

FIG. 6c Dependence of spin rotation angle $\phi$ as a function of particle's penetration depth into crystal $L$ for different initial antiproton angles with respect to crystal plane $\Theta = 0.25\,\Theta_L$ – solid line, $\Theta = 0.5\,\Theta_L$ – dashed line, $\Theta = 0.75\,\Theta_L$ – dotted line, $\gamma_0 = 10^6$, $\sigma_\gamma = 0.05\,\gamma_0$, $\sigma_\gamma = 0.05\Theta$.

Fig.6 shows the dependence of the averaged polarization vector components $\zeta_x$ (Fig. 6a), $\zeta_y$ (Fig. 6b), and spin rotation angle $\langle\phi\rangle$ (Fig. 6c) as a function of particle's penetration depth into crystal $L$ for different initial antiproton angles with respect to crystal plane $\Theta = 0.25\,\Theta_L$ – solid line, $\Theta = 0.5\,\Theta_L$ – dashed line, $\Theta = 0.75\,\Theta_L$ – dotted line. The energy and angular spreads are taken into account. The target is tungsten crystal oriented by plane (100); antiproton relativistic factor equals $\gamma_0=10^6$, $\sigma_\gamma = 0.05\,\gamma_0$, $\sigma_\gamma = 0.05\,\Theta$.

It follows from the Fig.6 that the energy and angular divergence in the antiproton beam leads to the damping of the amplitude of the oscillations of the polarization vector. Moreover, calculation demonstrates that the larger the spread, the greater the attenuation. In such a situation, the rotation of the particle's spin is determined mainly by the curvature of the crystal itself.

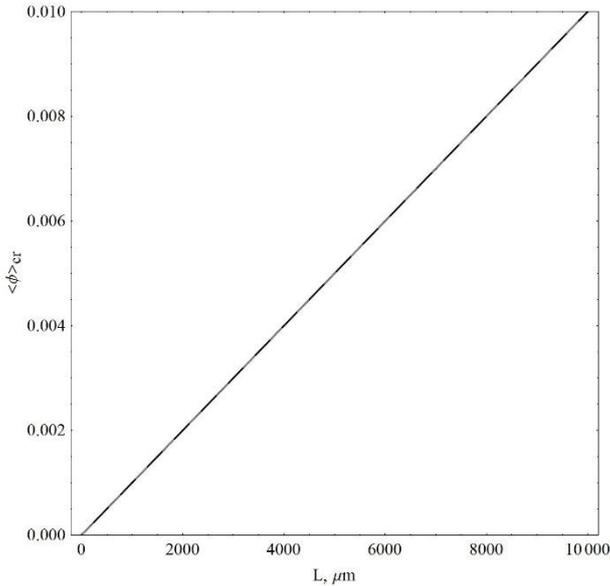

FIG. 8. Spin rotation angle $\langle\phi\rangle_{cr} = \langle\phi\rangle - \langle\phi\rangle_{st}$ due to crystal curvature.

In order to determinate the contribution of the crystal curvature into spin rotation angle $\langle\phi\rangle_{cr}$, in Fig. 8, we plot the difference between the rotation angles of the polarization vector in a straight $\langle\phi\rangle_{st}$ and curved crystal $\langle\phi\rangle$ ($\langle\phi\rangle_{cr} = \langle\phi\rangle - \langle\phi\rangle_{st}$ black line). For comparison, the same figure shows the angle of rotation calculated by the formula (17) (grey line). It follows from the figure that these results coincide with great accuracy. The target is tungsten crystal oriented by plane (100); the antiproton relativistic factor equals $\gamma=10^6$.

It is very important to note that these results practically independent of the angle of entry $\Theta$ of the particle and on whether or not we take into account the angular and energy spread in the antiproton beam.

The obtained numerical results are correct only for channeled particles. It means that the particle's penetration depth into the crystal should be much less than dechanneling length $L \ll l_d$. The dechanneling length can be estimated using the formula [47]

$$l_d = \frac{\alpha}{2\pi}\frac{2U_0\varepsilon}{m_e^2 c^4}L_{rad}.$$

Here $L_{rad}$ is the radiation length for an amorphous target and $\alpha$ is the fine structure constant, $m_e$ is the electron rest mass, $U_0$ is the depth of the potential well and $\varepsilon$ is the channeled particle total energy. For the conditions considered in the paper the dechanneling length of the order a few centimeters.

## V. CONCLUSION

We obtain an analytical solution to the Bargmann, Michel, and Teledgi equation describing the spin rotation of a negatively charged particle moving in a bent crystal in the planar channeling regime. If we discard the contribution from the bending of the crystal to the spin rotation angle in the obtained solution, then we obtain a solution for a straight crystal.

Our solution is valid for only a single trajectory of a particle in a crystal. Therefore, we average the solution over the possible trajectories of the channeled particles.

In real experiments, the antiprotons beam is characterized by energy and angular divergence and therefore the obtained results are averaged over the energy and angular dispersion. It is shown that the energy and angular spread in the antiproton beam lead to the fact that the oscillations of the polarization vector due to the periodic motion of particles in the crystal rapidly damp and the rotation of the particle's spin is determined mainly by the curvature of the crystal itself.

We did not take into account multiple scattering of particles in crystals, as well as the processes of dechanneling and capture of above-barrier particles in the channeling mode. Considering this should lead to even faster damping of the oscillations of the polarization vector due to the periodic trajectories of channeled particles in the crystal.

If we define, the angle of polarization vector rotation due to only crystal curvature as $\langle\phi\rangle_{cr} = \langle\phi\rangle - \langle\phi\rangle_{st}$, than it does not depend on the angle of entry $\Theta_0$ of the particle and also on whether or not we take into account the angular and energy spread in the antiproton beam. This angle $\langle\phi\rangle_{cr}$ is very well described by the formula (17) borrowed from the work of Luboschitz [2] if the angle of rotation of the particle is taken as the angle $\Psi = \Omega t$.

## ACKNOWLEDGMENTS

In conclusion, the author is sincerely grateful to X.T. Li for the fruitful discussions.